\documentclass[conference]{IEEEtran}
\IEEEoverridecommandlockouts

\usepackage{bbm}
\usepackage[utf8]{inputenc}
\usepackage{multirow}
\usepackage{colortbl}
\usepackage{booktabs}

\usepackage{cite}
\usepackage{amsmath,amssymb,amsfonts}
\usepackage{algorithm}
\usepackage{algpseudocode}
\algblock{ParFor}{EndParFor}
\usepackage{graphicx}
\usepackage{textcomp}
\usepackage{xcolor}
\def\BibTeX{{\rm B\kern-.05em{\sc i\kern-.025em b}\kern-.08em
    T\kern-.1667em\lower.7ex\hbox{E}\kern-.125emX}}

\algrenewcommand\algorithmicrequire{\textbf{Require:}}
\algrenewcommand\algorithmicensure{\textbf{Ensure:}}
\usepackage[caption=false,font=footnotesize]{subfig}

\def\BibTeX{{\rm B\kern-.05em{\sc i\kern-.025em b}\kern-.08em
    T\kern-.1667em\lower.7ex\hbox{E}\kern-.125emX}}
\begin{document}

\title{CroSatFL: Energy-Efficient Federated Learning with Cross-Aggregation for Satellite Edge Computing
}

\author{
\begin{tabular}{ccc}
\begin{tabular}{c}
Nan Yang\\
\textit{Western Sydney University}\\
n.yang@westernsydney.edu.au
\end{tabular}
&
\begin{tabular}{c}
Bahman Javadi\\
\textit{Western Sydney University}\\
b.javadi@westernsydney.edu.au
\end{tabular}
&
\begin{tabular}{c}
Rodrigo Neves Calheiros\\
\textit{Western Sydney University}\\
r.calheiros@westernsydney.edu.au
\end{tabular}
\\[0.5em]
\multicolumn{3}{c}{%
\begin{tabular}{cc}
\begin{tabular}{c}
David Boland\\
\textit{The University of Sydney}\\
david.boland@sydney.edu.au
\end{tabular}
&
\begin{tabular}{c}
Philip Leong\\
\textit{The University of Sydney}\\
philip.leong@sydney.edu.au
\end{tabular}
\end{tabular}%
}
\end{tabular}
}

\maketitle

\begin{abstract}
Low Earth Orbit (LEO) mega-constellations extend the cloud-to-edge continuum into space, enabling satellite edge computing. However, Federated Learning (FL) in this environment is fundamentally energy-constrained due to dynamic inter-satellite connectivity, heterogeneous onboard computing hardware, and strict power budgets.
We propose \emph{CroSatFL}, a sustainable on-orbit hierarchical FL framework that reduces end-to-end energy across computation and communication while maintaining strong training performance under realistic LEO dynamics. CroSatFL keeps the ground station (GS) off the iterative loop by performing all local training and intermediate aggregations on orbit, requiring only two GS communication phases: one for initialization and one for final model collection. This sharply reduces repeated use of bandwidth-limited and energy-expensive GS links and shifts iterative exchanges to laser inter-satellite links (LISLs). CroSatFL integrates three energy-aware mechanisms: \emph{StarMask} forms LISL-feasible clusters that align data volume with heterogeneous CPU/GPU capability, \emph{Skip-One} mitigates transient stragglers by skipping at most one slow client per cluster to lower round energy and latency while preserving long-term fairness, and \emph{random-$k$ cross-aggregation} enables lightweight topology-aware cross-cluster mixing without extending round duration. Using an end-to-end energy model with a realistic Walker-Delta constellation, we show that CroSatFL reduces GS communication count by over two orders of magnitude and GS transmission energy by about 6$\times$ relative to GS-centric and on-orbit baselines, while achieving competitive accuracy and faster convergence.
\end{abstract}


\section{Introduction}
Low Earth Orbit (LEO) mega-constellations such as Starlink and OneWeb are evolving into a planet-scale \emph{Satellite Edge Computing} substrate with continuous sensing, onboard processing, and low-latency laser inter-satellite links (LISLs) \cite{wang2023satellite}. This enables time-sensitive applications such as disaster response, environmental monitoring, maritime surveillance, and space situational awareness, where in-orbit learning reduces response latency and reliance on ground infrastructure. Nonetheless, downlinking remains a bottleneck due to short and misaligned ground station (GS) passes, limited satellite--ground bandwidth, and the high energy cost of long-range Radio Frequency (RF) transmission, making large-scale data transfer to Earth impractical. These constraints motivate pushing learning and decision-making into orbit.

Federated learning (FL) naturally fits this setting by shifting most training communication to on-orbit LISL exchanges \cite{fedsnTMC2025,fedleoTMC2024,jabbarpour2024fedorbit}. However, existing satellite and edge FL frameworks rely on abstractions that do not fully reflect LEO operations \cite{li2025advances}. Many remain GS-centric, coupling frequent global aggregation to short GS visibility windows and increasing latency and transmission energy on bandwidth-limited GS links \cite{fedleoTMC2024}. Representative examples include synchronous FedSyn \cite{mcmahan2017communication} and LEO-adapted FedLEO \cite{elmahallawy2023optimizing}. Even in on-orbit or hierarchical variants such as FELLO \cite{chen2023edge}, satellites are often treated as computationally uniform, overlooking runtime gaps across CPU/GPU and embedded accelerators \cite{bruhn2020enabling}. Energy-focused methods typically optimize isolated components, such as client selection in FedSCS \cite{al2025energy} or reduced-precision arithmetic in FedOrbit \cite{jabbarpour2024fedorbit}, without jointly addressing heterogeneous compute, time-varying LISL connectivity, and straggler-sensitive synchronization. As a result, clustering and scheduling under dynamic LISLs remain underexplored, and slow or poorly connected satellites can dominate round latency \cite{fedsnTMC2025}. These limitations motivate a fully on-orbit hierarchical FL framework that explicitly accounts for latency, communication cost, and energy under heterogeneous compute and time-varying LISLs.

We study a multi-orbital-plane LEO constellation where satellites train models on board and exchange FL updates over LISLs, using the GS only for bootstrap and final model collection. We focus on fully on-orbit hierarchical learning under dynamic, fan-out-limited LISLs \cite{cha2022survey} and strong CPU/GPU heterogeneity, where GS communication is significantly more expensive than short-range on-orbit transfer. The key challenge is to limit straggler-driven synchronization delay while bounding computation and communication energy under time-varying connectivity. To this end, we design \emph{CroSatFL} with three building blocks. \emph{StarMask} forms LISL-feasible, resource-aware clusters from runtime, hardware, and link profiles. \emph{Skip-One} reduces synchronization sensitivity by tolerating transient stragglers within each cluster while maintaining fairness over time. \emph{Random-$k$ cross-aggregation} enables lightweight model aggregation among reachable clusters based on instantaneous cross-plane connectivity. We summarize our contributions below.

\begin{itemize}
\item We propose \emph{CroSatFL}, a new hierarchical FL architecture that performs all intermediate aggregation on orbit and uses the ground station only for bootstrap and final model retrieval, avoiding GS-visibility bottlenecks and repeated high-cost satellite--ground communication.
\item We introduce \emph{StarMask}, a resource-aware clustering method that forms LISL-feasible and hardware-balanced clusters, and \emph{Skip-One}, a scheduling mechanism that alleviates synchronization bottlenecks by tolerating transient stragglers and consistently shortens round latency under heterogeneous compute.
\item We conduct an extensive performance evaluation to show that CroSatFL matches or exceeds existing GS-centric and on-orbit FL baselines in accuracy, while reducing GS communication count by over two orders of magnitude, reducing GS transmission energy by about 6$\times$, and significantly shortening end-to-end training time across datasets.
\end{itemize}

\section{Background and Related Work}
\label{sec:background}

\subsection{Intelligent Satellite Computing}
LEO constellations are evolving from passive relays into \emph{compute-capable} sensing systems that continuously collect imagery, RF traces, and multi-modal telemetry \cite{gardill2023towards}. Advances in processors, compact accelerators, and radiation-tolerant storage make in-orbit learning and inference increasingly feasible \cite{ortiz2023onboard}. Applications such as disaster response, environmental monitoring, maritime awareness, and space situational awareness require timely model updates and on-orbit decision support. Satellites can therefore preprocess and fine-tune models at the data source, while LISLs enable low-latency coordination to disseminate model progress across the constellation \cite{eagleeyeASPLOS2024,ai_spaceIEEE2023}. In practice, ground connectivity is intermittent and contention-prone, whereas LISL contacts are often sufficient to support on-orbit collaboration without relying on congested GS schedules \cite{kodanASPLOS2023,adaptiveCompressionTSC2024}. However, on-orbit learning remains constrained by tight and time-varying power, thermal, and memory/throughput budgets across platforms.

\subsection{Limitations of LEO Satellite Computing Systems}
\noindent\textbf{GS communication.} GS links are scarce, scheduled, and energy intensive, so routing intermediate model updates via ground stations makes the GS a synchronization bottleneck \cite{wirelessComm2022satFL,network2024satFL}. In practice, LEO satellites have only a few short contact windows per orbit day \cite{vasisht2021l2d2}, shared with payload downlink and other missions, so model exchange competes with higher-priority traffic and is vulnerable to visibility limits, congestion, and weather outages. In contrast, LISL-based on-orbit coordination is more frequent and can overlap with computation, enabling satellites to exchange updates with reachable peers while sensing continues. Yet many designs still centralize aggregation off orbit \cite{kodanASPLOS2023}, repeatedly paying uplink and downlink costs and tying training progress to non-learning factors such as handovers, rain fade, and GS contention.

\noindent\textbf{Hardware heterogeneity.}
LEO satellite constellations mix GPU-based and CPU-only satellites \cite{bruhn2020enabling}, creating large and time-varying gaps in per-epoch latency and energy due to thermal throttling and illumination-dependent power budgets. Under synchronous aggregation, round time is dictated by the slowest participants, inflating makespan and wasting energy as faster nodes wait at barriers. Earth-observation workloads further exacerbate this effect: cloudy, off-nadir, or night-pass imagery can slow training and yield weaker gradients, making straggling frequent and hard to predict \cite{fedleoTMC2024,fedsnTMC2025}. Yet many designs still use static heuristic clustering that ignores per-epoch time and energy and system constraints such as master buffering, LISL fan-out, and link budgets, leading to unbalanced groups that are difficult to synchronize \cite{batteryAwareTSC2024,eagleeyeASPLOS2024}. For time-critical surveillance or disaster response, this translates into wasted energy at aggregation nodes and delayed model refresh.

\subsection{Existing Solutions and Motivation}
Existing satellite FL solutions mainly coordinate training and aggregation under orbital dynamics and constrained communication. A representative line is GS-centric: satellites periodically upload updates to a GS for aggregation \cite{wirelessComm2022satFL,network2024satFL}. While simple, this assumes frequent and reliable GS contact, making it better suited to small constellations or low-rate updates.
To better exploit inter-satellite connectivity, recent work studies on-orbit or hierarchical aggregation over LISLs \cite{ai_spaceIEEE2023,tiansuanSDG2022}. These designs reduce GS dependence, but often rely on predefined or heuristic aggregation/participation policies, limiting adaptation to runtime variation and rapidly changing LISL topologies. Decentralized and offloading-assisted schemes leverage LISLs to reduce coordination overhead \cite{fedleoTMC2024,fedsnTMC2025}, while arithmetic simplification and adaptive compression reduce computation or communication cost \cite{jabbarpour2024fedorbit,adaptiveCompressionTSC2024}. Yet these techniques are typically used in isolation, offering limited end-to-end control of latency and energy at the constellation scale.

A key gap is that existing satellite FL systems do not jointly address the constraints that dominate LEO training in practice. Ground-centric designs assume frequent GS contact, while on-orbit variants often ignore dynamic LISLs fan-out limits, CPU/GPU heterogeneity, and straggler-induced waiting energy. As a result, they either place GS links on the critical path or waste energy and time at synchronization barriers, motivating a fully on-orbit, topology- and energy-aware design that remains accurate and fair under realistic LEO dynamics.

\section{System Model and Problem Formulation}
\subsection{System Model}
We consider a hierarchical FL system over a LEO satellite constellation with $N$ satellites
$\mathcal{S}=\{s_1,\ldots,s_N\}$. The on-orbit learning process is modeled as
\begin{equation}
\mathcal{F}
=
\big(\mathcal{S},\{\mathcal{D}_i\}_{i=1}^{N}, w^{(0)}, d, \mathcal{E}_{\mathrm{LISL}}(\cdot), \{c_i\}_{i=1}^N\big).
\label{eq:fl_system_def}
\end{equation}
Each satellite $s_i$ holds a private dataset $\mathcal{D}_i$ with $n_i=|\mathcal{D}_i|$ samples and trains
a model of size $d$ (bits) initialized by $w^{(0)}$. Inter-satellite communication is enabled by LISLs
with time-varying availability $\mathcal{E}_{\mathrm{LISL}}(t)$ and per-satellite fan-out limits $\{c_i\}$.
The LISL-reachable neighbors of $s_i$ at time $t$ are
$\mathcal{N}_i(t)=\{\,s_j \mid \{s_i,s_j\}\in\mathcal{E}_{\mathrm{LISL}}(t)\,\}$, satisfying
$|\mathcal{N}_i(t)| \le c_i$ for all $i$ and $t$.
The GS participates only at training boundaries to broadcast $w^{(0)}$ and retrieve the final model.
All intermediate aggregation and exchange occur on orbit, keeping the GS off the iterative critical path.

\noindent\textbf{Satellite Clustering.}
To exploit constellation structure and heterogeneity, we partition $\mathcal{S}$ into $K$ clusters
$\mathcal{C}=\{C_1,\ldots,C_K\}$ with $C_k\subseteq\mathcal{S}$ and $\bigcup_{k=1}^{K} C_k=\mathcal{S}$.
Within each cluster, intra-cluster aggregation is performed at a designated \emph{master satellite}.
Each satellite $s_i$ is characterized by $x_i=(n_i,h_i,T_i^{\mathrm{comp}},E_i^{\mathrm{train}},c_i)$,
capturing its data volume, hardware type, per-epoch computation time, per-round training energy, and
fan-out limit. Given the satellite profiles $\{x_i\}$ and the time-varying LISL topology
$\mathcal{E}_{\mathrm{LISL}}(t)$, we learn a LISL-feasible partition that balances connectivity stability,
compute capability, and resource availability. For each cluster $C_k$, the master $m_k(g)\in C_k$ is
selected dynamically in main round $g$ and may migrate across rounds based on orbital geometry,
instantaneous link quality, latency, and current load and resource conditions, while maintaining a
single logical aggregator per cluster.

\noindent\textbf{Hierarchical Federated Learning.}
After GS broadcast of $w^{(0)}$, training proceeds fully on orbit for $G$ \emph{main rounds} $g\in\{1,\ldots,G\}$ over $K$ clusters. In this work, we use a single main round ($G=1$). Within main round $g$, masters perform $R$ \emph{edge rounds} $r\in\{1,\ldots,R\}$ for iterative on-orbit training and random-$k$ cross-cluster exchanges over transient cross-plane LISLs. In each edge round $(g,r)$, cluster $C_k$ selects participants $P_k(r)\subseteq C_k$. Each $s_i\in P_k(r)$ runs $L_{\mathrm{loc}}$ \emph{local epochs} (one full pass over $\mathcal{D}_i$) and uploads its update to the current master $m_k(g)$, which aggregates updates to maintain the cluster model. If the master changes, the new master continues from the latest cluster model. In edge round $(g,r)$, cluster $k$ also derives the reachable neighbor set $\mathcal{N}_k^{\mathrm{reach}}(g,r)$ from the instantaneous LISL topology and samples a small subset for aggregation. After $R$ edge rounds, clusters are consolidated on orbit via sample-size weighted averaging, where $N_k=\sum_{i\in C_k} n_i$ denotes the data volume of cluster $k$. This keeps the GS off the iterative critical path.

\noindent\textbf{Satellite Participation Scheduling.}
Satellite participation is \emph{adaptive} rather than mandatory. In each edge round $r$, cluster $C_k$
selects $P_k(r)\subseteq C_k$ instead of enforcing full participation. Each satellite $s_i$ is assigned
an online utility score $U_i(r)$ that reflects its computation performance (e.g., $T_i^{\mathrm{comp}}$),
energy and memory state, communication opportunity under the instantaneous LISL topology and master
assignment, and historical contribution to training progress. Satellites with larger $U_i(r)$ are
prioritized for inclusion in $P_k(r)$, while those with smaller $U_i(r)$ may be skipped temporarily to reduce
straggling and avoid unnecessary energy use. The scores $U_i(r)$ are refreshed
so that a skipped satellite can rejoin once its resources recover or connectivity improves. This adaptive
scheduling reduces idle waiting, controls on-orbit energy consumption, and preserves long-term fairness by
avoiding systematic exclusion.

\subsection{Computation and Communication Model}\label{sec:comp-comm}
This section characterizes the computation and communication processes in hierarchical
federated learning over LEO constellations, capturing heterogeneous on-board computation,
time-varying inter-satellite connectivity, and GS communication constraints.

\noindent\textbf{Local Computation.}
One local epoch corresponds to a full pass over $\mathcal{D}_i$, and the per-epoch computation
workload scales linearly with data volume as
\begin{equation}
\mathrm{FLOPs}_i = n_i \cdot c_{\mathrm{flop}},
\label{eq:flops}
\end{equation}
where $c_{\mathrm{flop}}$ denotes the number of floating-point operations required per sample.
Let $T_i^{\mathrm{comp}}$ denote the per-epoch computation time of satellite $s_i$. Executing
$L_{\mathrm{loc}}$ local epochs in one training round incurs a total local training time
\begin{equation}
T_i^{\mathrm{train}} = L_{\mathrm{loc}} \cdot T_i^{\mathrm{comp}}.
\label{eq:local-train-time}
\end{equation}
To capture hardware heterogeneity, we introduce an effective computation throughput
$\alpha_i$ for each satellite and express the per-epoch runtime as
\begin{equation}
T_i^{\mathrm{comp}}=\frac{\mathrm{FLOPs}_i}{\alpha_i}.
\label{eq:comp-time}
\end{equation}
Satellites equipped with GPU-class accelerators typically exhibit significantly larger
$\alpha_i$ than CPU-only satellites. The parameter $\alpha_i$ provides a unified abstraction
of on-board computational capability. As a result, satellites with comparable dataset sizes
$n_i$ may experience substantially different training times, which motivates
heterogeneity-aware participation scheduling. After completing local training, satellites in
cluster $C_k$ transmit their model updates to the cluster master for aggregation. The
computational cost of aggregation scales as $O(|C_k|d)$ and is negligible compared to local training.

\noindent\textbf{Execution and Waiting Time.}
Due to intermittent GS visibility and time-varying LISL availability, a satellite
may be unable to immediately perform required communications after completing local
computation. We define the \emph{waiting time} as the wall-clock duration during which a
satellite performs no local training, aggregation, or transmission solely because required
GS or LISL communication opportunities are unavailable. Waiting time is treated as a
latency-only quantity and does not incur additional computation or transmission energy.

\noindent\textbf{LISL Communication.}
We adopt a unified communication model that treats all LISL transmissions as peer-to-peer
exchanges between satellites, regardless of whether the communicating nodes belong to the
same cluster or different orbital planes. The transmission delay for a model of size $d$ bits is
\begin{equation}
T^{\mathrm{LISL}}_{i\to j}(t)
=
\begin{cases}
\dfrac{d}{R^{\mathrm{LISL}}_{ij}(t)} + L^{\mathrm{LISL}}_{ij}(t), & \text{if } \{s_i,s_j\}\in\mathcal{E}_{\mathrm{LISL}}(t),\\[6pt]
\infty, & \text{otherwise},
\end{cases}
\label{eq:LISL_time}
\end{equation}
where $R^{\mathrm{LISL}}_{ij}(t)$ denotes the instantaneous link bandwidth and
$L^{\mathrm{LISL}}_{ij}(t)$ denotes the propagation latency between $s_i$ and $s_j$ at time $t$.
This model captures both high-bandwidth intra-plane LISLs used for uploading local updates
to cluster masters and intermittent cross-plane LISLs used for cross-cluster model exchange.
Variations in $R^{\mathrm{LISL}}_{ij}(t)$ and $L^{\mathrm{LISL}}_{ij}(t)$ reflect the
geometry-dependent dynamics of the constellation without requiring separate models for
different LISL types.

\noindent\textbf{Ground Station Communication.}
Communication between satellites and the ground station differs from LISLs due to visibility constraints and link characteristics. A satellite $s_i$ can communicate with the GS only when it enters the GS visibility window. Compared to LISLs, GS links generally exhibit lower bandwidth and higher latency. For a model of size $d$ bits, the
satellite--ground transmission delay is
\begin{equation}
T^{\mathrm{GS}}_i(t)=
\begin{cases}
\dfrac{d}{R^{\mathrm{GS}}_i(t)} + L^{\mathrm{GS}}_i(t), & \text{if $s_i$ is within GS coverage},\\[6pt]
\infty, & \text{otherwise},
\end{cases}
\label{eq:gs_time}
\end{equation}
where $R^{\mathrm{GS}}_i(t)$ and $L^{\mathrm{GS}}_i(t)$ denote the instantaneous bandwidth and
propagation delay of the satellite--ground link. In CroSatFL, GS communication is used only
for the initial model broadcast during session setup and a single final model downlink at
session completion.

\subsection{Energy Consumption Model}\label{sec:energy-model}
This subsection models satellite energy consumption in hierarchical FL.
We focus on local computation and communication energy under heterogeneous on-board
hardware and dynamic link conditions. During waiting periods, satellites are assumed to
operate in a low-power standby state without local training or GS and LISL transmissions.
Idle power is excluded to avoid conflating scheduling-induced latency with computation
efficiency.

\noindent\textbf{Training Energy Model.}
During local training, satellite $s_i$ consumes computation energy determined by its
hardware type and the number of samples processed in one round. Let
\begin{equation}
N_i = L_{\mathrm{loc}} \cdot n_i .
\label{eq:Ni_total_samples}
\end{equation}

For a \textbf{CPU-based satellite}, let $C_i^{\mathrm{CPU}}$ denote the number of CPU cycles
required per sample and $f_i^{\mathrm{CPU}}$ denote the operating frequency. Under a CMOS dynamic
power model with effective switched capacitance $\gamma_i$, the computation energy is
\begin{equation}
E_i^{\mathrm{CPU}}
=
\gamma_i \cdot C_i^{\mathrm{CPU}} \cdot N_i \cdot (f_i^{\mathrm{CPU}})^2 .
\label{eq:cpu_energy}
\end{equation}

For a \textbf{GPU-based satellite}, let $P_i^{\mathrm{avg}}$ denote the average power draw during
training. Using $T_i^{\mathrm{train}}$ from Eq.~\eqref{eq:local-train-time}, the corresponding
computation energy is
\begin{equation}
E_i^{\mathrm{GPU}}
=
P_i^{\mathrm{avg}} \cdot T_i^{\mathrm{train}}.
\label{eq:gpu_energy}
\end{equation}

We summarize the per-round computation energy of satellite $s_i$ as
\begin{equation}
E_i^{\mathrm{comp}} =
\begin{cases}
E_i^{\mathrm{CPU}}, & \text{if $s_i$ is CPU-based},\\
E_i^{\mathrm{GPU}}, & \text{if $s_i$ is GPU-based}.
\end{cases}
\label{eq:comp_energy}
\end{equation}
In this work, the per-round training energy is defined as
\begin{equation}
E_i^{\mathrm{train}} = E_i^{\mathrm{comp}} .
\label{eq:total_training_energy}
\end{equation}
We focus on computation energy, which dominates satellite power consumption during local
training and enables fair comparison across methods. Waiting periods incur no computation or transmission energy, since satellites are idle.

\noindent\textbf{LISLs Communication Energy Model.}
Satellites exchange model updates over time-varying LISLs whose availability, rate, and delay depend on orbital geometry. For a link between $s_i$ and $s_j$ at time $t$, let $R^{\mathrm{LISL}}_{ij}(t)$ denote the instantaneous data rate and $L^{\mathrm{LISL}}_{ij}(t)$ the propagation and processing delay. LISLs are used for both intra-cluster uploads to the master and intermittent inter-cluster exchanges among masters (random-$k$), each transmitting a payload of size $d$ bits.
The corresponding communication energy for one transmission is
\begin{equation}
E_{i\to j}^{\mathrm{LISL}}(t)
=
P_i^{\mathrm{LISL}} \cdot T^{\mathrm{LISL}}_{i\to j}(t),
\label{eq:LISL_energy}
\end{equation}
where $P_i^{\mathrm{LISL}}$ is the transmission power of $s_i$ and $T^{\mathrm{LISL}}_{i\to j}(t)$ follows Eq.~\ref{eq:LISL_time}. This model unifies intra- and inter-cluster LISLs communication and captures round-to-round energy variation induced by dynamic link conditions.
We model LISL energy at the transmission level and omit receiver-side and control-plane overhead, as these costs are comparable across methods and do not affect relative comparisons.

\begin{figure*}[t]
    \centering
    \includegraphics[width=0.9\textwidth]{Overview_Framework.pdf}
    \caption{Overview of the proposed framework for CroSatFL. StarMask first forms constraint-aware clusters over heterogeneous satellites. Skip-One then mitigates transient stragglers within each cluster, while random-$k$ cross-aggregation enables on-orbit model mixing among reachable masters. The ground station is used only for initialization and final model collection.}
    \label{fig:overview}
\end{figure*}

\noindent\textbf{GS Communication Energy Model.}
GS links are available only within visibility windows and typically require much higher transmission power than LISLs due to larger path loss and time-varying channel conditions. In CroSatFL, GS communication is invoked only at session boundaries, consisting of a one-time initial model broadcast and a single final model transfer for collection, and thus stays off the per-round critical path (Sec.~\ref{sec:comp-comm}). Let $R^{\mathrm{GS}}_i(t)$ and $L^{\mathrm{GS}}_i(t)$ denote the instantaneous GS rate and propagation/processing delay between the GS and satellite $s_i$ at time $t$. When $s_i$ is in coverage, the transmission time for a payload of size $d$ is $T^{\mathrm{GS}}_i(t)$ as in Eq. \eqref{eq:gs_time}. We model the GS communication energy for a single transfer as
\begin{equation}
E_i^{\mathrm{GS}}(t)
=
P^{\mathrm{GS}} \cdot T^{\mathrm{GS}}_i(t),
\label{eq:gs_energy}
\end{equation}
where $P^{\mathrm{GS}}$ is the effective transmission power associated with GS communication (including both GS-to-satellite and satellite-to-GS transfers). We model GS communication energy with an effective power that captures both uplink and downlink costs, and finer-grained distinctions do not affect our design choices or comparative results.

\section{Our Proposed Method: CroSatFL}\label{sec:CroSatFL}
CroSatFL is a fully on-orbit hierarchical FL framework for LEO constellations with time-varying LISLs and heterogeneous on-board resources. It integrates StarMask for
constraint-aware clustering, Skip-One for fair, edge-round-level straggler mitigation within each cluster, and random-$k$ cross-aggregation among cluster masters to maintain on-orbit global coherence as illustrated in Figure~\ref{fig:overview}. By keeping the ground station off the iterative critical path, CroSatFL reduces synchronization latency and computation energy.

\subsection{StarMask: RL-based Clustering with Action Masking}\label{sec:starmask}

\noindent\textbf{RL Formulation.}
Clustering is modeled as a finite-horizon Markov decision process (MDP) with horizon $N$, where
one satellite is assigned at each step. The decision maker is a parameterized \emph{policy
network} $\pi_\theta$ that maps the current state to a distribution over actions. The action
space has size $K_{\max}+1$, where $K_{\max}$ denotes the maximum allowable number of clusters.
Let
\begin{equation}
\mathrm{share}_i = \frac{n_i}{\sum_{j=1}^{N} n_j}
\end{equation}
denote the relative data share of satellite $s_i$.
At decision step $t$, the state is
\begin{equation}
s_t^{\mathrm{MDP}}
=
\big(x_t,\;\Phi(C_1),\ldots,\Phi(C_{K_{\max}})\big),
\label{eq:mdp-state}
\end{equation}
where $x_t=(\mathrm{share}_t,h_t,T_t^{\mathrm{comp}},E_t^{\mathrm{train}},c_t)$ summarizes the assigned satellite at step $t$, where $\mathrm{share}_t$ is the relative data share of the current satellite. It includes relative data share, hardware type, per-epoch runtime, per-round local training energy, and fan-out limit. The term $\Phi(C_k)$ summarizes cluster $C_k$ by its current size, per-epoch
time range, cumulative energy, data-share sum, hardware composition, and remaining capacity.
Uninstantiated clusters are treated as inactive until selected.

The action
\begin{equation}
a_t \in \{1,\ldots,K_{\max},K_{\max}+1\}
\end{equation}
assigns the current satellite to an existing cluster index in $\{1,\ldots,K_{\max}\}$ or selects
the $(K_{\max}+1)$-th action to open a new cluster. Transitions are deterministic and
incrementally update the partial partition. A terminal reward evaluates the completed
clustering $\mathcal{C}=\{C_1,\ldots,C_K\}$ as
\begin{equation}
\begin{aligned}
R(\mathcal{C})
= -\Big(
& \theta_{\mathrm{wait}} W(\mathcal{C})
+ \beta E_{\mathrm{tot}}(\mathcal{C}) \\
& + \gamma \sigma_{\mathrm{share}}^2(\mathcal{C})
+ \nu_{\#} K
+ \Lambda M_{\mathrm{mix}}(\mathcal{C})
\Big).
\end{aligned}
\end{equation}
where $K$ is the number of instantiated clusters. The term $W(\mathcal{C})$ measures the
intra-cluster per-epoch time mismatch by aggregating the per-cluster time ranges
\begin{equation}
W(\mathcal{C})=\sum_{k=1}^{K}\Big(\max_{i\in C_k} T_i^{\mathrm{comp}}-\min_{i\in C_k} T_i^{\mathrm{comp}}\Big),
\label{eq:W_def}
\end{equation}
and $E_{\mathrm{tot}}(\mathcal{C})$ is the per-epoch total cost that sums per-satellite
computation energy and intra-cluster LISL transmission energy under the resulting partition.
Let $\mathrm{share}(C_k)=\sum_{i\in C_k}\mathrm{share}_i$ denote the data share of cluster $C_k$.
We define the cross-cluster imbalance as
\begin{equation}
\sigma_{\mathrm{share}}^2(\mathcal{C})
=
\frac{1}{K}\sum_{k=1}^{K}\Big(\mathrm{share}(C_k)-\frac{1}{K}\sum_{\ell=1}^{K}\mathrm{share}(C_\ell)\Big)^2.
\label{eq:sigma_share_def}
\end{equation}
The term $M_{\mathrm{mix}}(\mathcal{C})$ penalizes heterogeneous hardware composition by counting
mixed-hardware clusters
\begin{equation}
M_{\mathrm{mix}}(\mathcal{C})
=
\sum_{k=1}^{K}\mathbbm{1}\!\left(\exists i,j\in C_k:\; h_i\neq h_j\right),
\label{eq:mix_def}
\end{equation}
where $\mathbbm{1}(\cdot)$ is the indicator function. The coefficients
$\theta_{\mathrm{wait}},\beta,\gamma,\nu_{\#},\Lambda$ are fixed across experiments to balance latency, energy, data-share fairness, the number of clusters, and hardware consistency. All reward terms are normalized using min--max ranges estimated from training instances so that no single component dominates the optimization.

\begin{algorithm}[t]
\caption{StarMask: RL-based Clustering}
\label{alg:starmask}
\footnotesize
\begin{algorithmic}[1]
\Require Ordered satellites $\mathcal{S}=\{s_1,\ldots,s_N\}$, profiles $\{x_i\}_{i=1}^N$,
maximum clusters $K_{\max}$, minimum cluster size $m_{\min}$, policy $\pi_\theta$
\Ensure Cluster partition $\mathcal{C}=\{C_1,\ldots,C_K\}$ or Infeasible with $K_{\min}$

\State Initialize empty clusters $\{C_1,\ldots,C_{K_{\max}}\}$; $K \gets 0$
\For{$t \gets 1$ to $N$}
    \State $s_t^{\mathrm{MDP}} \gets (x_t,\Phi(C_1),\ldots,\Phi(C_{K_{\max}}))$
    \Comment Eq.~\eqref{eq:mdp-state}
    \State $\mathcal{A}_t \gets \mathcal{A}(s_t^{\mathrm{MDP}})$
    \Comment Eq.~\eqref{eq:feasible-actions}
    \If{$\mathcal{A}_t = \emptyset$}
        \State Compute $K_{\min}$ using $\{\tilde{c}_\ell\}_{\ell=1}^N$
        \Comment Eq.~\eqref{eq:effective-capacity}
        \If{no feasible $K \in [K_{\min},K_{\max}]$ exists}
            \State \Return \textbf{Infeasible} with diagnostic $K_{\min}$
        \EndIf
        \State $\mathcal{C} \gets$ GreedyFallback$(\mathcal{S},K_{\max},m_{\min})$
        \State \Return $\mathcal{C}$
    \EndIf
    \State Sample $a_t \sim \pi_\theta(\cdot \mid s_t^{\mathrm{MDP}})$ restricted to $\mathcal{A}_t$
    \If{$a_t$ is \textsc{OpenNew}}
        \State Instantiate $C_{K+1}$ and assign the current satellite; $K \gets K+1$
    \Else
        \State Assign the current satellite to $C_{a_t}$
    \EndIf
\EndFor
\State \Return $\mathcal{C}=\{C_1,\ldots,C_K\}$
\end{algorithmic}
\end{algorithm}

The policy network $\pi_\theta$ is trained using an actor-critic method \cite{mnih2016asynchronous}
\begin{equation}
\nabla_\theta J(\theta)
=
\mathbb{E}_{\pi_\theta}
\!\left[
\sum_{t=1}^N 
\nabla_\theta \log \pi_\theta(a_t\,|\,s_t^{\mathrm{MDP}})\,A_t
\right],
\end{equation}
where $A_t$ denotes the advantage estimate at step $t$. The short horizon and terminal-only
rewards promote stable learning.

\noindent\textbf{Feasible Decision Making via Action Masking.}
To prevent infeasible intermediate assignments under strict LEO constraints, StarMask
enforces feasibility through action masking. From the raw action set
$\mathcal{A}_{\mathrm{raw}}$, feasible actions are defined as
\begin{equation}
\mathcal{A}(s_t^{\mathrm{MDP}})
=
\{a\in \mathcal{A}_{\mathrm{raw}}:\Gamma(s_t^{\mathrm{MDP}},a)=1\},
\label{eq:feasible-actions}
\end{equation}
where $\Gamma$ evaluates whether assigning the current satellite under action $a$ preserves all operational
constraints and allows completion of a feasible terminal partition.

For any cluster $C_k$, the master-feasibility constraint
\begin{equation}
|C_k|-1 \le \max_{j\in C_k} \tilde{c}_j
\end{equation}
is enforced, where $\tilde{c}_j$ denotes the effective fan-out capacity of satellite $s_j$.
When hardware-homogeneous clustering is required, assignments additionally satisfy
$h_t=h_j$ for all $j\in C_k$; otherwise, heterogeneity is penalized through $M_{\mathrm{mix}}$.
A minimum cluster size $m_{\min}$ is enforced at termination. During construction, $\Gamma$
also checks reachability based on remaining cluster capacity and unassigned satellites. The
``\textsc{OpenNew} cluster'' action is masked once $K=K_{\max}$.

\noindent\textbf{Attention-Guided Policy with Deterministic Fallback.}
StarMask employs attention to compute a relational embedding
\begin{equation}
z_t=\mathrm{Attn}(Q_t,K_t,V_t),
\end{equation}
where queries are derived from satellite features and keys and values from cluster summaries.
This embedding emphasizes per-epoch time compatibility, capacity headroom, and data-share
balance, enabling the policy network to select among feasible actions only.

\begin{algorithm}[t]
\caption{Skip-One Client Selection for Cluster $C_k$}
\label{alg:skip-one}
\footnotesize
\begin{minipage}{\columnwidth}
\begin{algorithmic}[1]
\Require Cluster $C_k$, local training times $\{T_i^{\mathrm{train}}\}_{i\in C_k}$,
computation energies $\{E_i^{\mathrm{train}}\}_{i\in C_k}$,
cooldown counters $\{\kappa_i(r)\}_{i\in C_k}$, staleness $\{\tau_i(r)\}_{i\in C_k}$
\Ensure Participant set $P_k(r)$ for edge round $r$

\State $S_k(r) \gets \emptyset$ \Comment initialize skipped set
\State Compute admissible skip set $\mathcal{U}_k(r)$ \Comment Eq.~\eqref{eq:admissible}

\If{$\mathcal{U}_k(r) = \emptyset$}
    \State $P_k(r) \gets C_k$
    \State \Return $P_k(r)$
\EndIf

\State Compute current barrier $M_k(r) \gets \max_{i\in C_k} T_i^{\mathrm{train}}$
\Comment Eq.~\eqref{eq:mk-def}

\ForAll{$i \in \mathcal{U}_k(r)$}
    \State $M_k^{(-i)}(r) \gets \max_{j\in C_k\setminus\{i\}} T_j^{\mathrm{train}}$
    \Comment Eq.~\eqref{eq:mk-counterfactual}
    \State $\Delta T_i(r) \gets M_k(r)-M_k^{(-i)}(r)$ \Comment Eq.~\eqref{eq:deltaT}
    \State $\Delta E_i^{\mathrm{train}}(r) \gets E_i^{\mathrm{train}}$ \Comment Eq.~\eqref{eq:deltaE-train}
    \State Compute utility $\Psi_k(\{i\};r)$ \Comment Eq.~\eqref{eq:utility}
\EndFor

\State $i^\star \gets \arg\max_{i\in\mathcal{U}_k(r)} \Psi_k(\{i\};r)$
\If{$\Psi_k(\{i^\star\};r) > 0$}
    \State $S_k(r) \gets \{i^\star\}$
    \State Update $\kappa_{i^\star}(r)$ and $\tau_{i^\star}(r)$
\EndIf

\State $P_k(r) \gets C_k \setminus S_k(r)$ \Comment Eq.~\eqref{eq:pk-def}
\State \Return $P_k(r)$
\end{algorithmic}
\end{minipage}
\end{algorithm}

If no feasible action exists, StarMask invokes a deterministic fallback. It computes a lower
bound on the number of required clusters using effective capacities
\begin{equation}
\tilde{c}_i=\min(c_i-1,L_{h_i}),
\label{eq:effective-capacity}
\end{equation}
where $L_{h_i}$ denotes the hardware-dependent upper bound on the manageable number of member satellites for a master of type $h_i$. StarMask then greedily constructs the smallest feasible partition by assigning satellites in descending per-epoch runtime order while enforcing fan-out, hardware, and minimum-size constraints. The
resulting partition satisfies $K_{\min}\le K\le K_{\max}$. If no feasible partition exists,
StarMask reports infeasibility together with $K_{\min}$. The overall StarMask workflow is
summarized in Algorithm~\ref{alg:starmask}.

\subsection{Skip-One Client Selection}\label{sec:skip-one}
Skip-One Client Selection is a lightweight, edge-round-level mechanism within each cluster
$C_k$. In this work, all edge rounds are executed within a single main round ($G=1$). Despite long-term clustering, short-term variations in orbital geometry, LISL quality, and compute load can induce transient stragglers. To reduce synchronization barriers, the master $m_k$ may skip at most one satellite per edge round $r$, following Algorithm~\ref{alg:skip-one}, to balance latency reduction, computation-energy saving, and long-term participation fairness. A skipped satellite performs no local training and sends no update to the master in that edge round.
This design is intentionally conservative: it targets the common case where a single transient straggler dominates the cluster barrier, often the main source of delay in our relatively small clusters, while avoiding overly aggressive exclusion that could harm participation fairness. Extending the rule to skip multiple satellites is possible, but is beyond the scope of this work.

\noindent\textbf{Barrier reduction under transient stragglers.}
Let the realized participant set in round $r$ be
\begin{equation}
P_k(r) = C_k \setminus S_k(r), \qquad |S_k(r)|\le 1,
\label{eq:pk-def}
\end{equation}
where $S_k(r)$ denotes the skipped satellite set. The cluster-level computation barrier before skipping is defined as
\begin{equation}
M_k(r) = \max_{i\in C_k} T_i^{\mathrm{train}},
\label{eq:mk-def}
\end{equation}
where $T_i^{\mathrm{train}}$ denotes the local training time of satellite $s_i$. Communication
delays over LISLs are handled separately and are not included in the computation barrier.

To evaluate the effect of skipping a candidate satellite $i\in C_k$, we consider a
counterfactual participant set $C_k \setminus \{i\}$, yielding the counterfactual barrier
\begin{equation}
M_k^{(-i)}(r) = \max_{j\in C_k\setminus\{i\}} T_j^{\mathrm{train}} \le M_k(r),
\label{eq:mk-counterfactual}
\end{equation}
and the resulting latency reduction
\begin{equation}
\Delta T_i(r) = M_k(r) - M_k^{(-i)}(r) \ge 0.
\label{eq:deltaT}
\end{equation}
This formulation captures the common LEO scenario where a satellite temporarily becomes a
straggler despite being well matched by the long-term cluster assignment.

\noindent\textbf{Computation-energy saving.}
Skipping satellite $i$ in round $r$ excludes it from local training and eliminates its
computation energy consumption for that round. The resulting energy saving is
\begin{equation}
\Delta E_i^{\mathrm{train}}(r) = E_i^{\mathrm{train}},
\label{eq:deltaE-train}
\end{equation}
where $E_i^{\mathrm{train}}$ denotes the per-round computation energy of satellite $s_i$.
Waiting-induced idle energy is not considered. Skip-One therefore reduces round latency
through barrier tightening while explicitly accounting for computation-energy savings.

\noindent\textbf{Fairness-aware admissible set.}
To prevent systematic exclusion of specific satellites, Skip-One restricts candidates to
the admissible skip set
\begin{equation}
\mathcal{U}_k(r)=
\{ i\in C_k:\; \kappa_i(r)=0,\ \tau_i(r)<\tau_{\max} \},
\label{eq:admissible}
\end{equation}
where $\kappa_i(r)$ is a cooldown counter preventing consecutive skips, and $\tau_i(r)$
measures participation staleness. These constraints ensure bounded exclusion duration and
preserve long-term participation fairness.

\noindent\textbf{Round-level selection rule.}
The cluster master selects the skip decision as
\begin{equation}
S_k(r)\in
\arg\max_{S\subseteq\mathcal{U}_k(r),\,|S|\le 1}
\Psi_k(S;r),
\label{eq:skip-argmax}
\end{equation}
with $\Psi_k(\emptyset;r)=0$ and
\begin{equation}
\Psi_k(\{i\};r)
=
\theta_T\,\Delta T_i(r)
+\theta_E\,\Delta E_i^{\mathrm{train}}(r)
-\theta_H\,\mathcal{H}_i
-\theta_F\,\phi_i(r),
\label{eq:utility}
\end{equation}
where $\mathcal{H}_i$ is a static hardware-aware penalty that discourages skipping satellites
with rare or high-value hardware configurations, and $\phi_i(r)$ encodes recent participation
history. The weights $\theta_T,\theta_E,\theta_H,\theta_F$ are fixed across experiments, and
all terms are normalized to comparable ranges to balance latency reduction, energy saving,
and fairness. The resulting selection incurs $O(|\mathcal{U}_k(r)|)$ per-round complexity and
relies only on locally available statistics.

Since $\Delta T_i(r)\ge 0$, Skip-One weakly reduces the cluster computation barrier $M_k(r)$
and improves round latency. Periodic all-participation rounds reset cooldown counters,
ensuring that all satellites contribute regularly over long horizons despite transient
straggler events.

\subsection{Cross-Aggregation without GS Communication}\label{sec:cross-agg}
CroSatFL maintains global model coherence fully on orbit via direct LISL-based exchanges among cluster masters. Since cross-plane LISLs are intermittent and the master graph is rarely connected, CroSatFL uses lightweight \emph{random-$k$ cross-aggregation}, where each master mixes with a small set of currently reachable neighbors in each edge round.

\begin{figure*}[t]
    \centering
    \begin{minipage}{0.3\textwidth}
        \centering
        \includegraphics[width=\textwidth]{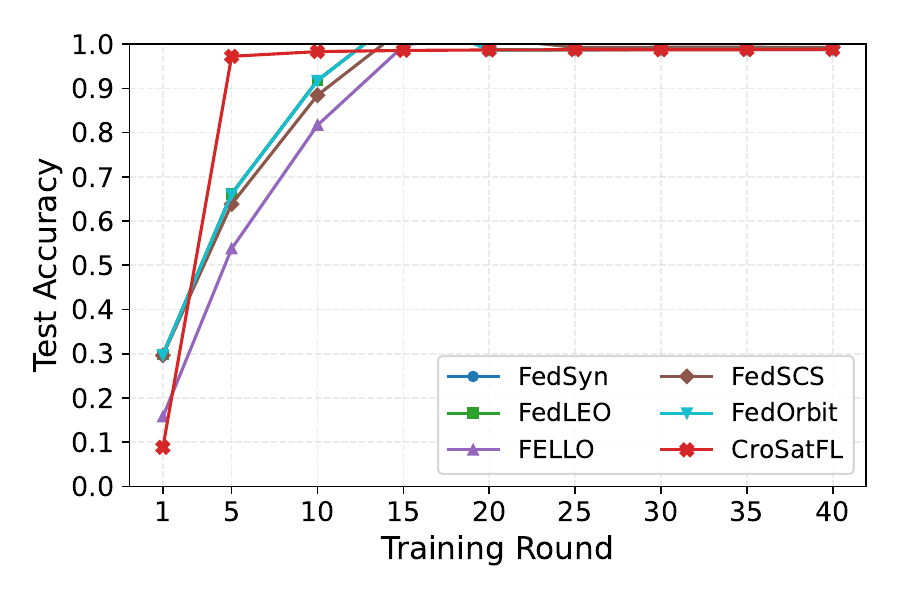}
        \vspace{-1mm}
        {\footnotesize Accuracy for ResNet-18-MNIST (IID)}
    \end{minipage}
    \hfill
    \begin{minipage}{0.3\textwidth}
        \centering
        \includegraphics[width=\textwidth]{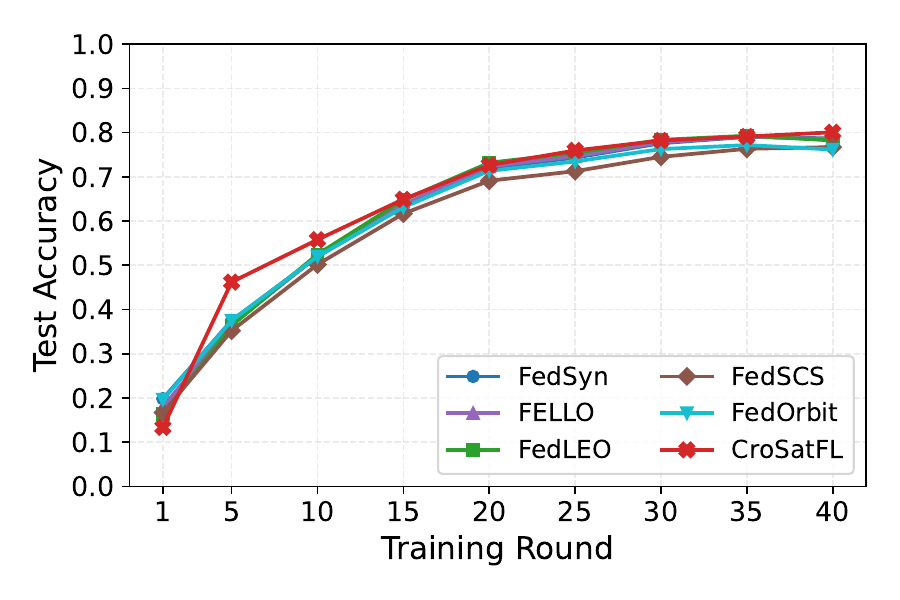}
        \vspace{-1mm}
        {\footnotesize Accuracy for ResNet-18-CIFAR-10 (IID)}
    \end{minipage}
    \hfill
    \begin{minipage}{0.3\textwidth}
        \centering
        \includegraphics[width=\textwidth]{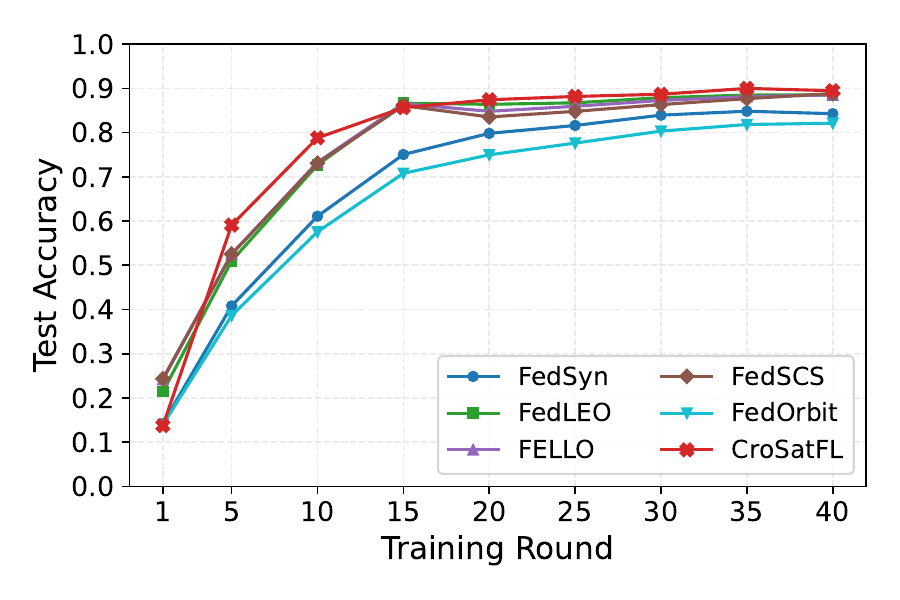}
        \vspace{-1mm}
        {\footnotesize Accuracy for ResNet-18-EuroSAT (IID)}
    \end{minipage}

    \caption{Convergence comparison on MNIST, CIFAR-10, and EuroSAT using ResNet-18 under IID settings.}
    \label{fig:IID}
\end{figure*}

\begin{figure*}[t]
    \centering
    \begin{minipage}{0.32\textwidth}
        \centering
        \includegraphics[width=\textwidth]{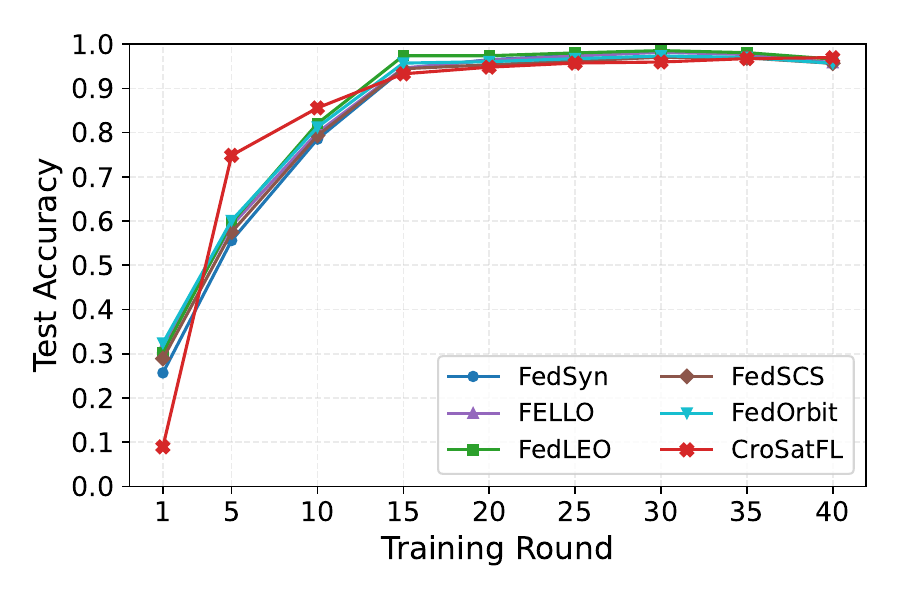}
        \vspace{-1mm}
        {\footnotesize Accuracy for ResNet-18-MNIST (non-IID)}
    \end{minipage}
    \hfill
    \begin{minipage}{0.32\textwidth}
        \centering
        \includegraphics[width=\textwidth]{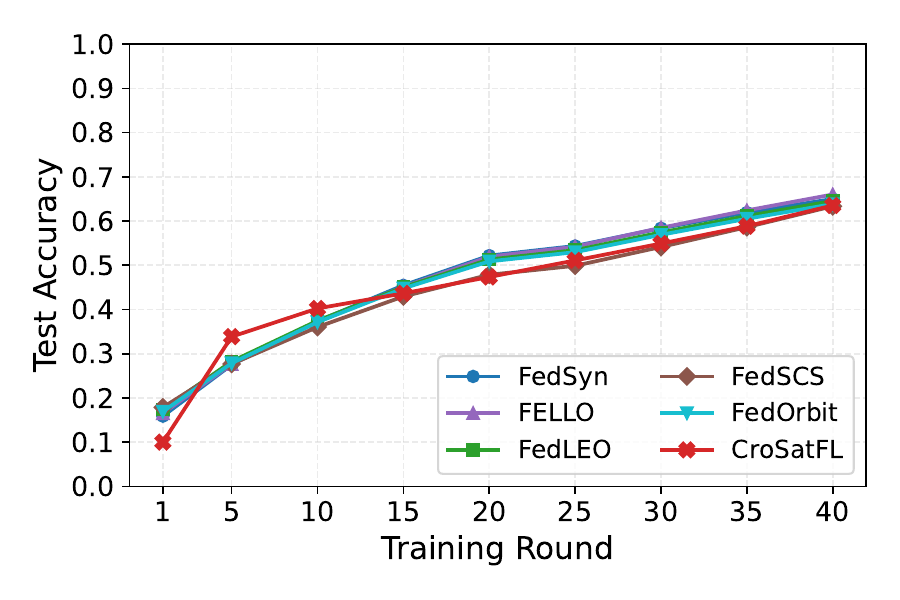}
        \vspace{-1mm}
        {\footnotesize Accuracy for ResNet-18-CIFAR-10 (non-IID)}
    \end{minipage}
    \hfill
    \begin{minipage}{0.32\textwidth}
        \centering
        \includegraphics[width=\textwidth]{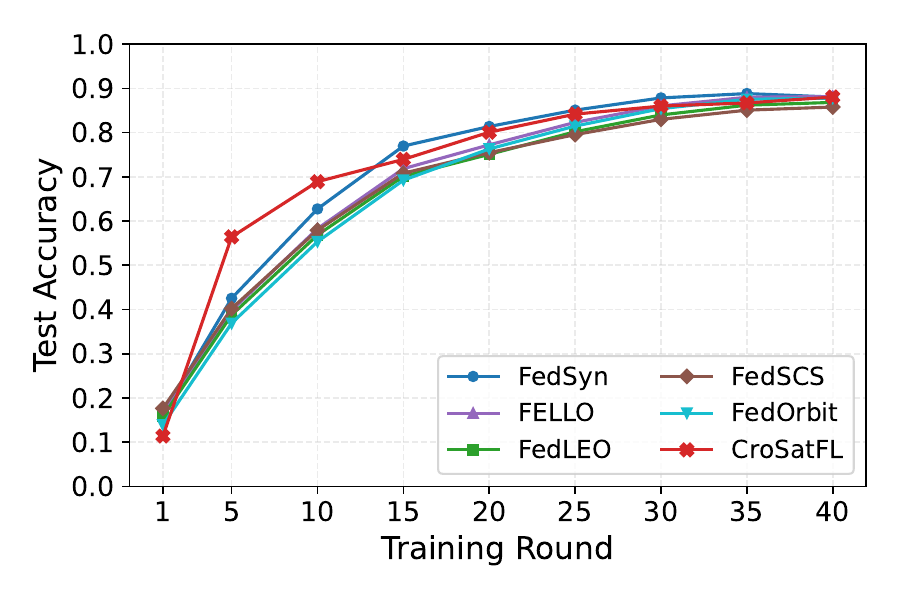}
        \vspace{-1mm}
        {\footnotesize Accuracy for ResNet-18-EuroSAT (non-IID)}
    \end{minipage}

    \caption{Convergence comparison on MNIST, CIFAR-10, and EuroSAT using ResNet-18 under non-IID settings.}
    \label{fig:NonIID}
\end{figure*}

\noindent\textbf{Random-$k$ LISL-based cross-aggregation.}
Let $w_k^{g}$ denote the model held by the master of cluster $C_k$ after completing the
intra-cluster aggregation stage of main round $g$. We initialize the edge-round model by setting
$w_k^{g,1}=w_k^{g}$. Let
\begin{equation}
N_k=\sum_{i\in C_k} n_i
\label{eq:Nk_cross}
\end{equation}
denote the total number of samples hosted by cluster $C_k$. Within main round $g$, cluster
masters perform $R$ edge rounds indexed by $r\in\{1,\ldots,R\}$. In edge round $(g,r)$,
cluster $k$ observes the instantaneous LISL topology and identifies the set of reachable
clusters $\mathcal{N}_k^{\mathrm{reach}}(g,r)$. From this set, it uniformly samples a
neighbor subset
\begin{equation}
\begin{aligned}
\mathcal{N}_k(g,r)&\subseteq\mathcal{N}_k^{\mathrm{reach}}(g,r), \\
|\mathcal{N}_k(g,r)| &= \min\!\left(k_{\mathrm{nbr}},\,|\mathcal{N}_k^{\mathrm{reach}}(g,r)|\right).
\end{aligned}
\label{eq:neighbor-set}
\end{equation}
where $k_{\mathrm{nbr}}$ is a fixed sampling parameter. This definition keeps the sampling
rule well-defined even when fewer than $k_{\mathrm{nbr}}$ neighbors are reachable. The local
mixing group is then defined as
\begin{equation}
\mathcal{M}_k(g,r) = \{k\}\cup \mathcal{N}_k(g,r).
\label{eq:mixing-group}
\end{equation}
The master of cluster $C_k$ updates its model by performing a sample-size weighted average
\begin{equation}
w_k^{g,r+1}
=
\sum_{j\in\mathcal{M}_k(g,r)}
\frac{N_j}{\sum_{\ell\in\mathcal{M}_k(g,r)} N_\ell}\, w_j^{g,r},
\label{eq:randomk-mixing}
\end{equation}
where $w_j^{g,r}$ denotes the model of cluster $j$ available at the beginning of edge round
$r$. Uniform neighbor sampling preserves statistical proportionality through
data-weighted aggregation. Over successive edge rounds, the random-$k$ exchanges induced
by orbital motion propagate information across orbital planes, yielding a gossip-like
consensus process at the cluster level without requiring full connectivity at any single time.

\noindent\textbf{On-orbit consolidation and once-per-session GS communication.}
After completing the $R$ edge rounds in the final main round $G$, each cluster $C_k$ holds a
model $w_k^{G,R+1}$ that has incorporated information from multiple neighboring clusters. The
constellation then forms the final global model entirely on orbit as
\begin{equation}
w^{(\mathrm{final})}
=
\sum_{k=1}^{K}
\frac{N_k}{\sum_{j=1}^{K} N_j}\, w_k^{G,R+1},
\label{eq:final-model}
\end{equation}
using sample-size weighted averaging across clusters. After completing all $G$ main rounds, the
final on-orbit model is downlinked to the ground station for checkpointing or deployment. This
design avoids placing GS communication on the round-level critical path and eliminates
dependence on aligned GS visibility windows.

Together, random-$k$ cross-aggregation enables topology-aware and statistically correct global
mixing while avoiding additional round-level synchronization. Combined with StarMask clustering
and Skip-One client selection, it completes the CroSatFL design for fully on-orbit,
energy-efficient hierarchical federated learning under realistic LEO dynamics.

\begin{table}[t]
\centering
\caption{Experimental Parameters}
\label{tab:parameters}
\begin{tabular}{l c}
\hline
\textbf{Parameter} & \textbf{Value} \\
\hline
Constellation & Walker Delta (Starlink) \\
Number of LEO satellites & 720 \\
Number of orbits & 36 \\
Number of LEOs per Orbit & 20 \\
Inclination & $70^\circ$ \\
Altitude & 570 km \\
Allocated spectrum Bandwidth of ISL/GS & 2.5/1.25 GHz \\
System loss $(h)$ & 3 dB \\
Noise power $(\sigma)$ & $2.2 \times 10^{-16}$ W \\
Frequency & 27 GHz \\
Gain to noise temperature ratio $(G)$ & 5 dB/K \\
Data rate & 16 Mbps \\
Transmission power $(p)$ & 40 Watts \\
FL main rounds & 1 \\
FL edge rounds & 40 \\
Local training epochs & 10 \\
Batch size & 10 \\
\hline
\end{tabular}
\end{table}

\begin{figure*}[t]
    \centering
    \begin{minipage}{0.3\textwidth}
        \centering
        \includegraphics[width=\textwidth]{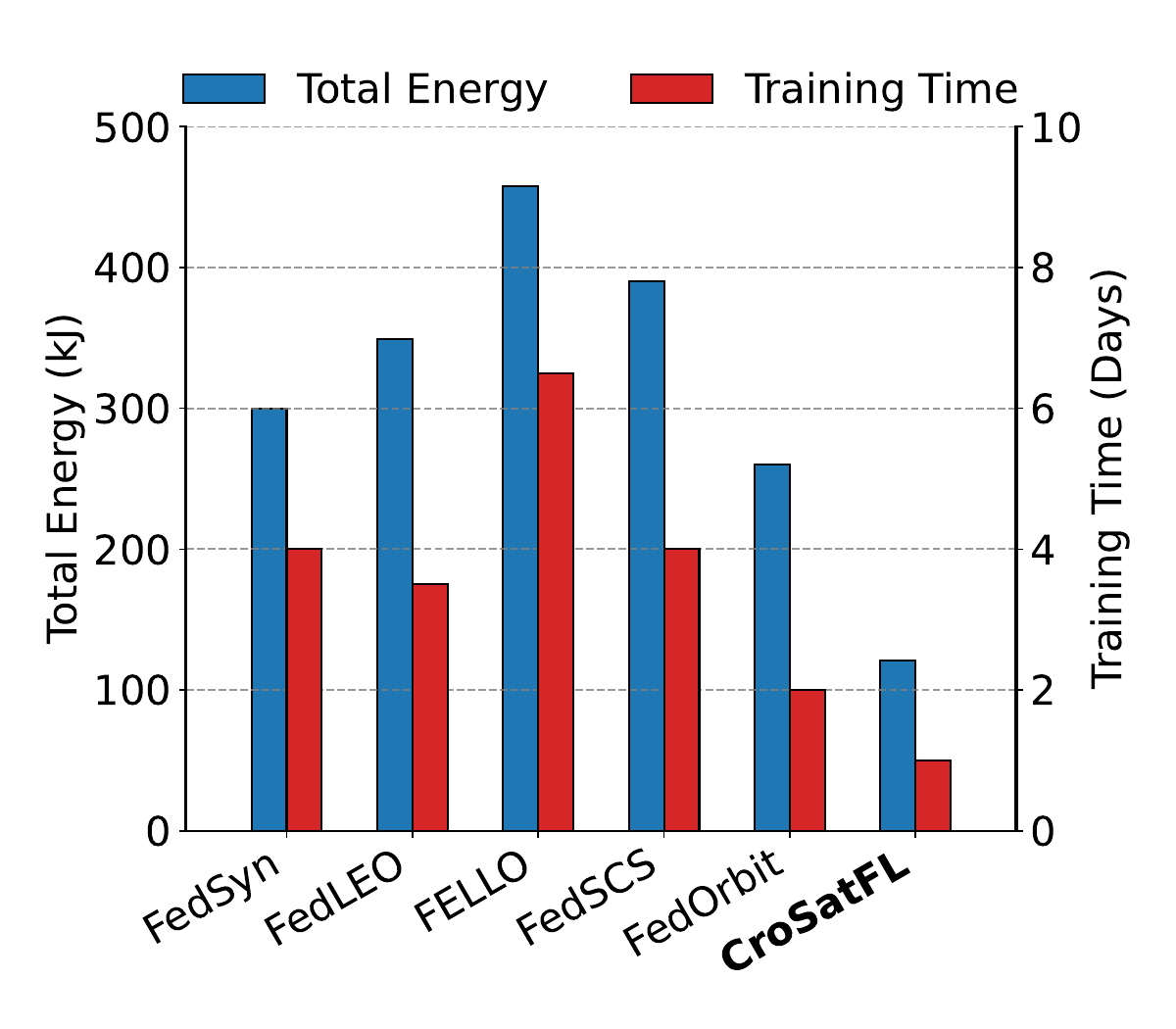}
        \vspace{-1mm}
        {\footnotesize ResNet-18-MNIST (Acc. 95\%)}
    \end{minipage}
    \hfill
    \begin{minipage}{0.3\textwidth}
        \centering
        \includegraphics[width=\textwidth]{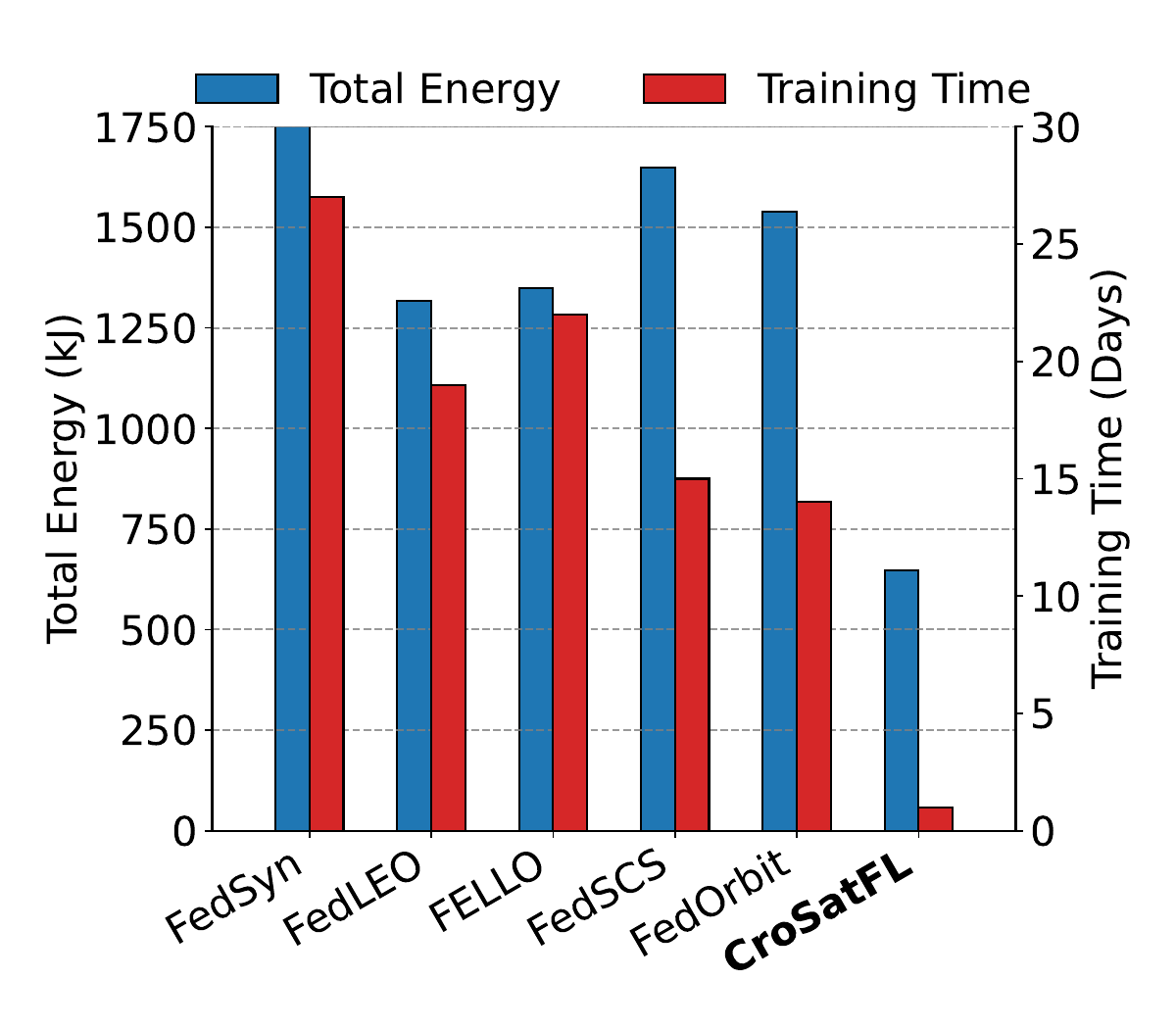}
        \vspace{-1mm}
        {\footnotesize ResNet-18-CIFAR-10 (Acc. 75\%)}
    \end{minipage}
    \hfill
    \begin{minipage}{0.3\textwidth}
        \centering
        \includegraphics[width=\textwidth]{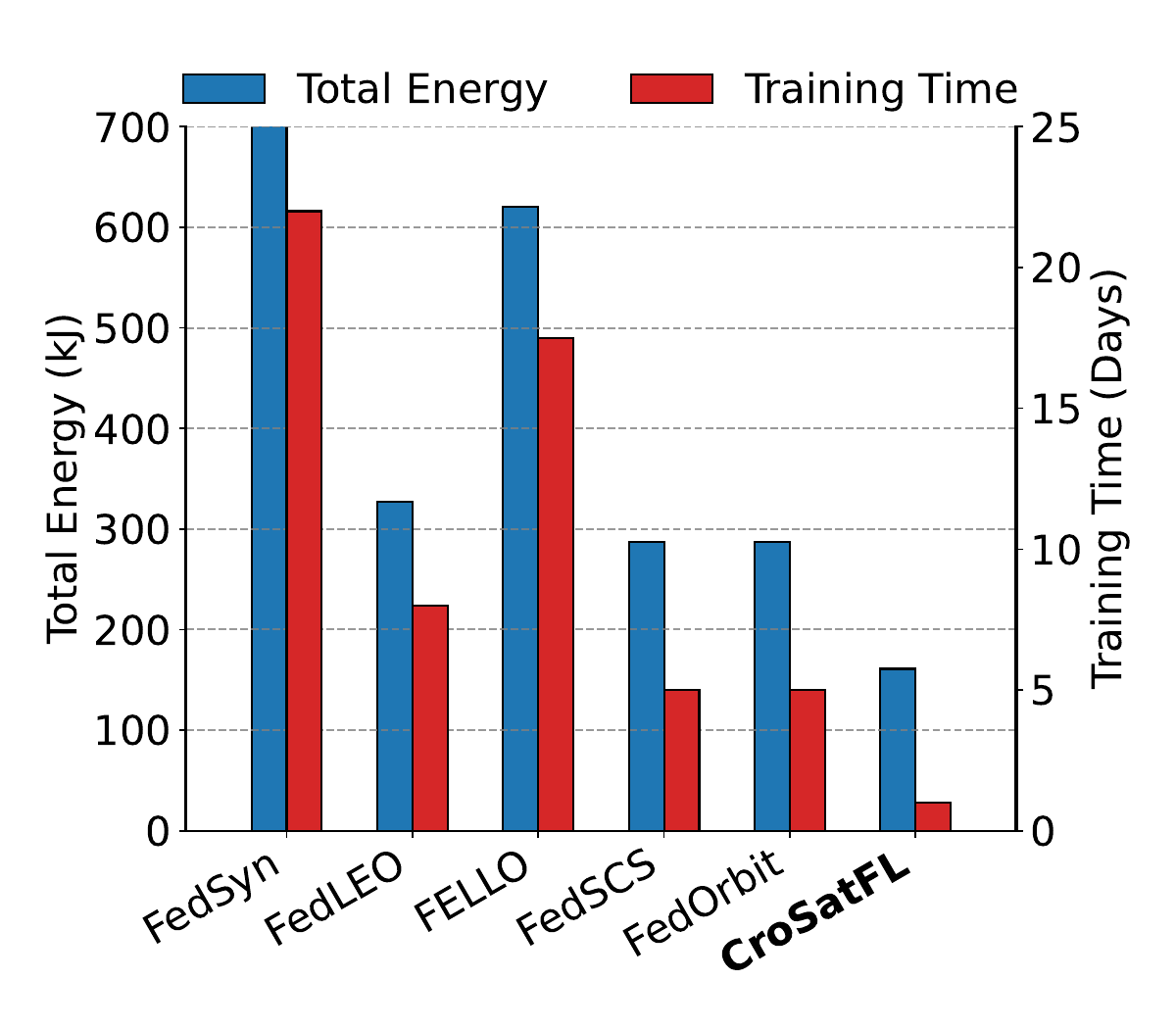}
        \vspace{-1mm}
        {\footnotesize ResNet-18-EuroSAT (Acc. 80\%)}
    \end{minipage}

    \caption{Comparison of total energy consumption and end-to-end training time for a common target accuracy.}
    \label{fig:three_energy}
\end{figure*}

\section{Experimental Results}
We evaluate CroSatFL on multiple datasets using model accuracy, energy consumption, and training time. CroSatFL is implemented in Flower \cite{beutel2020flower} with PyTorch and executed on  $\text{NVIDIA\textregistered}$ RTX 5090 GPU. 
Satellite computing capacities follow real Spiral Blue Space Edge One\footnote{https://www.spiralblue.space/} traces collected in 2023 from multiple in-orbit tests. These traces are used under a simulated constellation in which 50\% of satellites are CPU-only and 50\% are GPU-equipped.

\subsection{Experimental Setup}
We consider a Walker-Delta constellation of 720 LEO satellites at 570~km altitude and $70^\circ$ inclination, arranged in 36 orbital planes with 20 satellites per plane (inter-/intra-plane spacings $10^\circ$/$18^\circ$), instantiated using the MATLAB Satellite Communications Toolbox. We place a ground station in Canberra, Australia (latitude $-35.40139^\circ$, longitude $148.98167^\circ$). We consider four LISL communication-range settings, i.e., 659, 1319, 1500, and 1700 km, which approximately support maximum cluster sizes of 2, 4, 6, and 10, respectively. We follow system parameters from prior work \cite{jabbarpour2024fedorbit} as listed in Table~\ref{tab:parameters}. We randomly select 40 satellites as FL clients and use StarMask to form 9 clusters, each managed by a master.  

\noindent\textbf{Baselines.} 
We compare CroSatFL with five representative baselines, each implemented following its
original design, including its native communication and synchronization mechanisms.
Baselines are not constrained to adopt CroSatFL's once-per-session GS communication or
on-orbit cross-aggregation.
\emph{FedSyn}~\cite{mcmahan2017communication} is standard synchronous FedAvg.
\emph{FedLEO}~\cite{elmahallawy2023optimizing} adapts FL to LEO via intra-plane
propagation and sink-satellite scheduling. \emph{FELLO}~\cite{chen2023edge} uses
optical-LISL clustering and edge selection to reduce communication overhead.
\emph{FedSCS}~\cite{al2025energy} performs energy-aware client selection for orbital
edge computing. \emph{FedOrbit}~\cite{jabbarpour2024fedorbit} reduces computation and
communication cost with block minifloat arithmetic. Together, they cover synchronous FL,
clustering-based methods, topology-aware scheduling, and energy-efficient orbital FL.

\begin{table}[t]
    \centering
    \caption{Breakdown of LISL/GS communication, energy consumption, and waiting time for EuroSAT}
    \label{tab:comm_energy}
    \scriptsize
    \setlength{\tabcolsep}{3pt}
    \renewcommand{\arraystretch}{1.2}
    \begin{tabular}{lcccccc}
        \hline
        & FedSyn & FELLO & FedLEO & FedSCS & FedOrbit & \textbf{CroSatFL }\\
        \hline
        Intra-cluster LISLs & -- & 3120 & 2800 & 2560 & 2560 & 1760 \\
        (No.) & & & & & & \\
        Inter-cluster LISLs & -- & -- & 0 & 0 & 0 & 1440 \\
        (No.) & & & & & & \\
        GS Communication & 3200 & 80 & 400 & 640 & 640 & 18 \\
        (No.) & & & & & & \\
        Transmission Energy & 601.60 & 108.90 & 159.48 & 197.38 & 197.38 & 99.70 \\
        Cost (kJ) & & & & & & \\
        Training Energy & 1080 & 1080 & 1080 & 1080 & 1080 & 179.18 \\
        Cost (kJ) & & & & & & \\
        Transmission Time & 0.87 & 0.20 & 0.27 & 0.32 & 0.32 & 0.19 \\
        (Hours) & & & & & & \\
        Waiting Time & 936.25 & 816.92 & 696.85 & 456.80 & 456.80 & 7.89 \\
        (Hours) & & & & & & \\
        \hline
    \end{tabular}
\end{table}

\noindent\textbf{Model and Datasets.} We evaluate our framework using ResNet-18 \cite{he2016deep} on three standard image classification datasets: MNIST \cite{lecun2002gradient}, CIFAR-10 \cite{krizhevsky2009learning}, and EuroSAT \cite{helber2019eurosat} for both IID and non-IID data distributions.

\subsection{Experimental Results}
\noindent\textbf{Model Accuracy.} 
Figures \ref{fig:IID} and \ref{fig:NonIID} compare the convergence of CroSatFL and baselines under IID and non-IID data. Under IID settings, CroSatFL achieves the best final accuracy on CIFAR-10 (80.09\%) and a top-performing result on EuroSAT (89.49\%), slightly outperforming or matching all baselines. Under non-IID ($\alpha=0.5$), where heterogeneity slows and destabilizes training for most baselines, CroSatFL maintains robust learning dynamics and achieves competitive final performance across all datasets, remaining close to the strongest methods. Overall, these results show that the proposed CroSatFL preserves convergence stability and learning effectiveness across both homogeneous and heterogeneous data scenarios, supporting scalable satellite FL.

\noindent\textbf{Energy Consumption.} 
Figure~\ref{fig:three_energy} compares total energy and training time for six FL frameworks to reach the same target accuracy on MNIST (95\%), CIFAR-10 (75\%), and EuroSAT (80\%) with ResNet-18. Across all datasets, CroSatFL achieves the lowest energy consumption and shortest training time. FedSyn appears relatively low because MNIST is simple and ResNet-18 converges quickly, but under lighter models and more realistic settings, it remains the slowest and most energy-hungry baseline \cite{al2025energy}. Table~\ref{tab:comm_energy} shows that CroSatFL shifts communication to inter-cluster LISLs and sharply reduces costly GS interactions, cutting GS communications from 3200 to 18 and transmission energy from 601.60 kJ to 99.70 kJ. It also reduces straggler-induced synchronization delay, shrinking waiting time from hundreds of hours to 7.89 hours and lowering end-to-end latency. These results reveal that CroSatFL improves scalability by reducing both communication burden and synchronization-driven delays.

\begin{figure}[t]
    \centering
    \includegraphics[width=0.85\columnwidth]{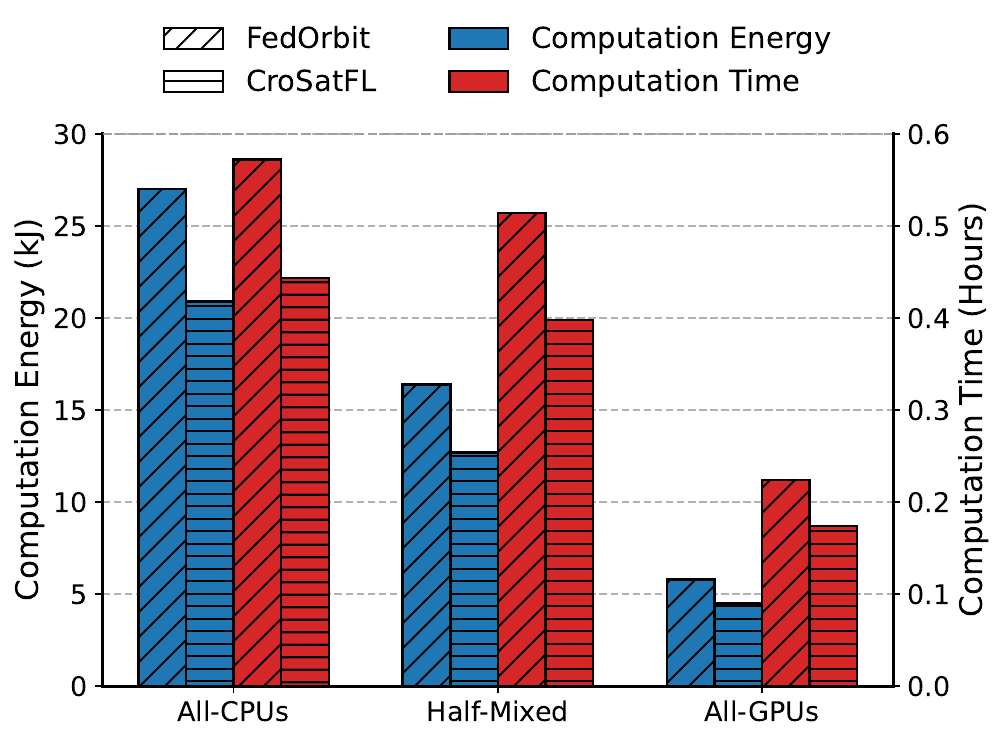}
    \caption{Single edge round computation energy and time comparison under heterogeneous CPU/GPU settings}
    \label{fig:cpugpu}
\end{figure}

\noindent\textbf{Impact of Heterogeneous Hardware Composition.}
Figure~\ref{fig:cpugpu} compares one-round computation energy and training time of CroSatFL and FedOrbit under three hardware compositions: All-CPUs (100\% CPUs), Half-Mixed (50\% CPUs/50\% GPUs), and All-GPUs (100\% GPUs). FedOrbit is used as the baseline because it explicitly considers heterogeneous on-board hardware (i.e., FPGA) and thus serves as a representative reference. With full participation, its round time is dominated by the slowest satellite and is therefore more sensitive to hardware heterogeneity. In contrast, CroSatFL adopts skip-one scheduling to alleviate straggler effects and better utilize GPU-equipped satellites. As the GPU share increases, both methods achieve lower energy consumption and latency, while CroSatFL consistently maintains lower one-round cost and stronger robustness across different hardware mixes.

\section{Conclusion}
This paper presents \emph{CroSatFL}, a fully on-orbit hierarchical federated learning framework that eliminates frequent ground station dependence and enables energy-efficient training under realistic LEO constraints. CroSatFL keeps all intermediate aggregation in space and uses the ground station only for session bootstrap and final model retrieval, removing GS visibility from the critical path and avoiding costly, misaligned contacts. To address dynamic, fan-out-limited LISL connectivity and strong CPU/GPU heterogeneity, CroSatFL integrates \emph{StarMask} for LISL-feasible, resource-balanced clustering, \emph{Skip-One} for straggler-resilient round control, and \emph{random-$k$} cross-aggregation for topology-aware inter-cluster mixing without increasing round latency. Experiments show that CroSatFL preserves competitive accuracy while reducing GS communication count by over two orders of magnitude, reducing GS transmission energy by about 6$\times$, and consistently shortening end-to-end training time. Future work will explore adaptive compression and fault-tolerant scheduling, and validate CroSatFL under richer orbital dynamics and onboard accelerators.

\section*{Acknowledgment}
We thank the Space Research Network and the NSW State Government for financial support of this project through the SRN Pilot Project grant 2024-25.

\bibliographystyle{IEEEtran}
\bibliography{references}

@article{fedleoTMC2024,
  author = {Z. Zhai and Q. Wu and S. Yu and R. Li and F. Zhang and X. Chen},
  title = {FedLEO: An Offloading-Assisted Decentralized Federated Learning Framework for Low Earth Orbit Satellite Networks},
  journal = {IEEE Transactions on Mobile Computing},
  year = {2024}
}

@article{fedsnTMC2025,
  author = {Z. Lin and Z. Chen and Z. Fang and X. Chen and X. Wang and Y. Gao},
  title = {FedSN: A Federated Learning Framework over Heterogeneous LEO Satellite Networks},
  journal = {IEEE Transactions on Mobile Computing},
  year = {2025}
}

@article{wirelessComm2022satFL,
  author = {H. Chen and M. Xiao and Z. Pang},
  title = {Satellite-Based Computing Networks with Federated Learning},
  journal = {IEEE Wireless Communications},
  year = {2022}
}

@article{network2024satFL,
  author = {B. Matthiesen and N. Razmi and I. Leyva-Mayorga and A. Dekorsy and P. Popovski},
  title = {Federated Learning in Satellite Constellations},
  journal = {IEEE Network},
  year = {2024}
}

@article{batteryAwareTSC2024,
  author = {Li, Qing and Wang, Shangguang and Ma, Xiao and Zhou, Ao and Wang, Yue and Huang, Gang and Liu, Xuanzhe},
  title = {Battery-aware energy optimization for satellite edge computing},
  journal = {IEEE Transactions on Services Computing},
  year = {2024},
  publisher = {IEEE}
}

@inproceedings{kodanASPLOS2023,
  author = {B. Denby and K. Chintalapudi and R. Chandra and B. Lucia and S. Noghabi},
  title = {Kodan: Addressing the Computational Bottleneck in Space},
  booktitle = {ASPLOS},
  year = {2023}
}

@article{adaptiveCompressionTSC2024,
  author = {C. Yang and others},
  title = {Towards Efficient Satellite Computing Through Adaptive Compression},
  journal = {IEEE Transactions on Services Computing},
  year = {2024}
}

@inproceedings{eagleeyeASPLOS2024,
  author = {Z. Cheng and B. Denby and K. McCleary and B. Lucia},
  title = {EagleEye: Nanosatellite Constellation Design for High-Coverage, High-Resolution Sensing},
  booktitle = {ASPLOS},
  year = {2024}
}

@article{ai_spaceIEEE2023,
  author = {X. Xu and Q. Wang and Y. Hou and S. Wang},
  title = {{AI-SPACE}: A Cloud-Edge Aggregated Artificial Intelligent Architecture for Tiansuan Constellation-Assisted Space-Terrestrial Integrated Networks},
  journal = {IEEE Network},
  year = {2023}
}

@incollection{tiansuanSDG2022,
  author = {Y. Tang},
  title = {Tiansuan Constellation: Intelligent Software-Defined Microsatellite with Orbital Attention for Sustainable Development Goals},
  booktitle = {ICBDIC},
  publisher = {Springer},
  year = {2022}
}

@article{wang2023satellite,
  author = {Shangguang Wang and Qing Li},
  title = {Satellite Computing: Vision and Challenges},
  journal = {IEEE Internet of Things Journal},
  year = {2023}
}

@article{jabbarpour2024fedorbit,
  author = {Mohammad Reza Jabbarpour and Bahman Javadi and Philip H. W. Leong and Rodrigo N. Calheiros and David Boland},
  title = {FedOrbit: Energy Efficient Federated Learning for Orbital Edge Computing Using Block Minifloat Arithmetic},
  journal = {IEEE Transactions on Services Computing},
  year = {2024}
}

@article{beutel2020flower,
  author = {Beutel, Daniel J and Topal, Taner and Mathur, Akhil and Qiu, Xinchi and Fernandez-Marques, Javier and Gao, Yan and Sani, Lorenzo and Li, Kwing Hei and Parcollet, Titouan and De Gusm{\~a}o, Pedro Porto Buarque and others},
  title = {Flower: A friendly federated learning research framework},
  journal = {arXiv preprint arXiv:2007.14390},
  year = {2020}
}

@inproceedings{mcmahan2017communication,
  author = {McMahan, Brendan and Moore, Eider and Ramage, Daniel and Hampson, Seth and y Arcas, Blaise Aguera},
  title = {Communication-efficient learning of deep networks from decentralized data},
  booktitle = {AISTATS},
  year = {2017},
  organization = {PMLR}
}

@inproceedings{elmahallawy2023optimizing,
  author = {Elmahallawy, Mohamed and Luo, Tie},
  title = {Optimizing federated learning in LEO satellite constellations via intra-plane model propagation and sink satellite scheduling},
  booktitle = {IEEE ICC},
  year = {2023},
  organization = {IEEE}
}

@inproceedings{chen2023edge,
  author = {Chen, Chih-Yu and Shen, Li-Hsiang and Feng, Kai-Ten and Yang, Lie-Liang and Wu, Jen-Ming},
  title = {Edge selection and clustering for federated learning in optical inter-LEO satellite constellation},
  booktitle = {PIMRC},
  year = {2023},
  organization = {IEEE}
}

@inproceedings{al2025energy,
  author = {Al-Blewi, Bara’ah and Javadi, Bahman and Calheiros, Rodrigo N},
  title = {Energy Efficient Client Selection in Federated Learning for Orbital Edge Computing},
  booktitle = {IEEE EDGE},
  year = {2025},
  organization = {IEEE}
}

@article{lecun2002gradient,
  author = {LeCun, Yann and Bottou, L{\'e}on and Bengio, Yoshua and Haffner, Patrick},
  title = {Gradient-based learning applied to document recognition},
  journal = {Proceedings of the IEEE},
  year = {2002},
  publisher = {Ieee}
}

@misc{krizhevsky2009learning,
  author = {Krizhevsky, Alex and Hinton, Geoffrey and others},
  title = {Learning multiple layers of features from tiny images},
  year = {2009},
  publisher = {Toronto, ON, Canada}
}

@article{helber2019eurosat,
  author = {Helber, Patrick and Bischke, Benjamin and Dengel, Andreas and Borth, Damian},
  title = {Eurosat: A novel dataset and deep learning benchmark for land use and land cover classification},
  journal = {IEEE Journal of Selected Topics in Applied Earth Observations and Remote Sensing},
  year = {2019},
  publisher = {IEEE}
}

@inproceedings{he2016deep,
  author = {He, Kaiming and Zhang, Xiangyu and Ren, Shaoqing and Sun, Jian},
  title = {Deep residual learning for image recognition},
  booktitle = {CVPR},
  year = {2016}
}

@inproceedings{li2025advances,
  author = {Li, Zilinghan and He, Shilan and Yang, Ze and Ryu, Minseok and Kim, Kibaek and Madduri, Ravi},
  title = {Advances in appfl: A comprehensive and extensible federated learning framework},
  booktitle = {CCGrid},
  year = {2025},
  organization = {IEEE}
}

@article{cha2022survey,
  author = {Cha, Hong Seol and Kim, Jong Min and Lim, Byungju and Lee, Ju Hyung and Ko, Young Chai},
  title = {A survey on inter-satellite links for low-earth orbit satellite networks},
  journal = {Journal of Korean Institute of Communications and Information Sciences},
  year = {2022},
  publisher = {Korean Institute of Communications and Information Sciences}
}

@misc{gardill2023towards,
  author = {Gardill, Markus and Kinser, Witold and Budroweit, Jan and Al-Hourani, Akram and Dunkel, Emily and Swope, Jason and Evans, David and Staebler, Maximilian},
  title = {Towards space edge computing and onboard AI for real-time teleoperations},
  year = {2023},
  publisher = {IEEE CS Press}
}

@article{bruhn2020enabling,
  author = {Bruhn, Fredrik C and Tsog, Nandinbaatar and Kunkel, Fabian and Flordal, Oskar and Troxel, Ian},
  title = {Enabling radiation tolerant heterogeneous GPU-based onboard data processing in space},
  journal = {CEAS Space Journal},
  year = {2020},
  publisher = {Springer}
}

@article{ortiz2023onboard,
  author = {Ortiz, Flor and Monzon Baeza, Victor and Garces-Socarras, Luis M and V{\'a}squez-Peralvo, Juan A and Gonzalez, Jorge L and Fontanesi, Gianluca and Lagunas, Eva and Querol, Jorge and Chatzinotas, Symeon},
  title = {Onboard processing in satellite communications using ai accelerators},
  journal = {Aerospace},
  year = {2023},
  publisher = {MDPI}
}

@inproceedings{vasisht2021l2d2,
  author = {Vasisht, Deepak and Shenoy, Jayanth and Chandra, Ranveer},
  title = {L2D2: Low latency distributed downlink for LEO satellites},
  booktitle = {SIGCOMM},
  year = {2021}
}

@inproceedings{mnih2016asynchronous,
  author = {Mnih, Volodymyr and Badia, Adria Puigdomenech and Mirza, Mehdi and Graves, Alex and Lillicrap, Timothy and Harley, Tim and Silver, David and Kavukcuoglu, Koray},
  title = {Asynchronous methods for deep reinforcement learning},
  booktitle = {ICML},
  year = {2016},
  organization = {PMLR}
}

\end{document}